\documentclass[pdftex]{nic-series}
\usepackage{macros} 
\begin{document}
%
%
\title{%
{\vspace*{-2.75cm}\small\hfill\textnormal{%
MS-TP-10-03
}}\\[1.75cm]
Towards Precision B-physics from Non-Perturbative\\
Heavy Quark Effective Theory
}

\author{%
ALPHA Collaboration\\
Michele Della Morte \inst{1} 
\and
Jochen Heitger \inst{2}
}

\institute{%
Johannes Gutenberg Universit\"at Mainz, Institut f\"ur Kernphysik,\\
Johann-Joachim-Becher Weg~45, D-55099 Mainz, Germany\\
\email{morte@kph.uni-mainz.de}
\and
Westf\"alische Wilhelms-Universit\"at M\"unster,
Institut f\"ur Theoretische Physik,\\ 
Wilhelm-Klemm-Stra{\ss}e~9, D-48149 M\"unster, Germany\\
\email{heitger@uni-muenster.de}
}

\maketitle

\begin{abstracts}
We convey an idea of the significant recent progress, which opens up good 
perspectives for high-precision ab-initio computations in heavy flavour 
physics based on lattice QCD.
Rather than surveying the latest results, this contribution focuses on the 
concept and the challenges of fully non-perturbative computations in the 
B-meson sector, where the b-quark is treated within an effective theory.
We outline its use to determine the b-quark mass and report on the results
obtained in the quenched approximation and on the status in the two 
dynamical flavour theory.
\end{abstracts}

\section{B-physics and lattice QCD}
\label{Sec_Bphys}
The plenty of beautiful results from recent and still ongoing B-physics 
experiments~\cite{HFAG,Altarelli:2009ec}, which require the knowledge of QCD 
matrix elements for their interpretation in terms of parameters of the 
Standard Model and its possible extensions, motivates investigations in 
lattice QCD.
The importance of this interplay of experiment and theory is further
expressed by the fact that one of its main objectives, the phenomenon of CP 
violation, is closely related to the symmetry breaking mechanism
the 2008's Nobel Prize was dedicated to.

Lattice QCD represents our best founded theoretical formulation of QCD and 
allows for the computation of low-energy hadronic properties in the 
non-perturbative domain, where the usual power series expansion in the 
coupling constant fails, through the Monte Carlo evaluation of the Euclidean 
path integral after a discretization of space-time on a lattice with spacing 
$a$ in all 3$+$1 dimensions~\cite{nic01:rainer+harmut,reviews:NPRrainer_nara}.
While such numerical computations necessarily involve approximations,
one of the key features of the lattice approach is that all approximations 
can be systematically improved. 
For an overview of results from the field of heavy flavour physics, which
reflect some of these improvements by the small error bars quoted for many 
quantities, we refer to the reviews of past Lattice 
Conferences~\cite{lat07:michele,lat08:bphys,lat09:bphys}.
\subsection{Challenges}
\label{Sec_Bphys_chall}
\begin{figure}[htb]
\begin{center}
\vspace{-1.1875cm}
\includegraphics[width=0.775\textwidth]{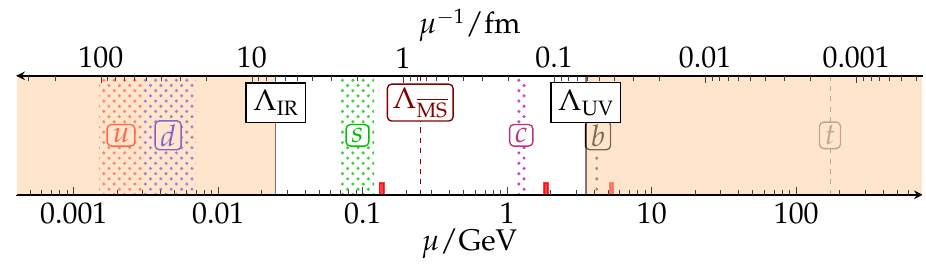}
\vspace{-0.5cm}
\caption{%
Large range of energy ($\mu$) scales in lattice QCD, where shaded areas 
refer to quark mass values (in the $\MSbar$ scheme) quoted by the Particle 
Data Group \cite{PDBook}.
Red marks indicate the pion, the D- and the B-meson mass. 
}\label{fig:massscales}
\end{center}
\vspace{-0.5cm}
\end{figure}

Among the various considerable challenges one faces in an actual lattice
QCD calculation on the theoretical and technical levels, let us only 
highlight the multi-scale problem, which is also particularly relevant in 
view of B-physics applications.
This is illustrated in \Fig{fig:massscales}.
%
%
There are many disparate physical scales to be covered simultaneously, 
ranging from the lightest hadron mass of $\mpi\approx 140\,\MeV$ over
$\mD\approx 2\,\GeV$ to $\mB\approx5\,\GeV$, plus the ultraviolet cutoff of 
$\Lambda_{\rm UV}=a^{-1}$ of the lattice discretization that has to be large 
compared to all physical energy scales for the discretized theory to be an 
approximation to the continuum one.
Moreover, the finiteness of the linear extent of space-time, $L$, in a 
numerical treatment entails an infrared cutoff $\Lambda_{\rm IR}=L^{-1}$ so 
that the following scale hierarchy is met:
\beq
\Lambda_{\rm IR}\,=\,L^{-1}
\quad\ll\quad 
\mpi\,,\,\ldots\,,\,\mD\,,\,\mB
\quad\ll\quad
a^{-1}\,=\,\Lambda_{\rm UV}\,. 
\eeq
This implies $L\gtrsim 4/\mpi\approx 6\,\Fm$ to suppress finite-size effects 
in the light quark sector and $a\lesssim 1/(2\mD)\approx 0.05\,\Fm$ to still
properly resolve the propagation of a c-quark in the heavy sector.
Lattices with $L/a \gtrsim 120$ sites in each direction would thus be 
needed to satisfy these constraints, and since the scale of hadrons with
b-quarks was not even included to arrive at this figure, it is obvious that 
the b-quark mass scale has to be separated from the others in a 
theoretically sound way before simulating the theory.
In \Sect{Sec_npHQET} we describe, how this is achieved by recoursing to an 
effective theory for the b-quark.

Another non-trivial task is the renormalization of QCD operators composed 
of quark and gluon fields, which appear in the effective weak Hamiltonian, 
valid at energies far below the electroweak scale.
Besides perturbation theory \cite{LPT:stefano}, powerful non-perturbative 
approaches have been 
developed \cite{nic01:rainer+harmut,reviews:NPRrainer_nara}, and we will
come back to the non-perturbative subtraction of power-law divergences in
the context of the effective theory for the b-quark later.
\subsection{Perspectives}
\label{Sec_Bphys_persp}
As for the challenges with light quarks, we only mention that the condition 
$L\gtrsim 6\,\Fm$ may be relaxed by simulating at unphysically large 
pion masses, combined with a subsequent extrapolation guided by chiral 
perturbation theory~\cite{chPT:GaLe1} and its lattice-specific refinements.

Regarding the algorithmic side of a lattice QCD simulation, the Hybrid Monte 
Carlo \cite{hmc:orig} (HMC) as the first exact and still state-of-the-art 
algorithm has received considerable improvements by multiple time-scale 
integration schemes \cite{hmc:mtsi1,hmc:mtsi2}, the Hasenbusch trick of 
mass-preconditioning \cite{hmc:hasenb1,hmc:hasenb2}, supplemented by a
sensible tuning of the algorithm's parameters \cite{Nf2SF:algo}, and the
method of domain decomposition (DD) applied to 
QCD \cite{ddhmc:luescher1,ddhmc:luescher2,ddhmc:luescher3}, just to name
a few.
In addition, low-mode deflation \cite{deflat:luescher2} (together with
chronological inverters \cite{Brower:1995vx}) has led to a substantial 
reduction of the critical slowing down with the quark mass in the DD-HMC.

Finally, in parallel to the continuous increase of computer speed
(at an exponential rate) over the last 25 years and the recent investments 
into high performance computing at many places of the world, the
Coordinated Lattice Simulations \cite{community:CLS} (CLS) initiative is a 
community effort to bring together the human and computer resources of 
several teams in Europe interested in lattice QCD.
The present goal are large-volume simulations with $\nf=2$ dynamical quarks,
using the rather simple $\Or(a)$ improved Wilson 
action~\cite{reviews:NPRrainer_nara} to profit from the above algorithmic 
developments such as DD-HMC, and lattice spacings $a=0.08,0.06,0.04\,\Fm$, 
sizes $L=(2-4)\,\Fm$ and pion masses down to $\mpi=200\,\MeV$, which 
altogether help to diminish systematic and statistical errors.
Amongst others, charm physics~\cite{lat08:mcharmNf2} as well as our 
B-physics programme outlined here are being investigated. 

\section{Non-perturbative Heavy Quark Effective Theory}
\label{Sec_npHQET}
Heavy Quark Effective Theory (HQET) at zero velocity on the 
lattice~\cite{stat:eichhill1} offers a reliable solution to the problem of 
dealing with the two disparate intrinsic scales encountered in heavy-light 
systems involving the b-quark, i.e., the lattice spacing $a$, which has to 
be much smaller than $1/m_{\rm b}$ to allow for a fine enough resolution of 
the states in question, and the linear extent $L$ of the lattice volume, 
which has to be large enough for finite-size effects to be under control
(recall also \Fig{fig:massscales}).

Since the heavy quark mass ($\mb$) is much larger than the other scales 
such as its 3--momentum or $\lQCD\sim 500\,\MeV$, HQET relies upon a 
systematic expansion of the QCD action and correlation functions in inverse 
powers of the heavy quark mass around the static limit ($\mb\to\infty$).
The lattice HQET action at $\Or(1/\mb)$ reads:
\beq
S_{\rm HQET}=
a^4{\T \sum_x}\heavyb\left\{
D_{0}+\dmstat-\omkin\vecD^2-\omspin\vecsigma\vecB
\right\}\heavy\,,
\eeq
with $\heavy$ satisfying $P_+\heavy=\heavy$, $P_+={{1+\gamma_0}\over{2}}$,
and the parameters $\omega_{\rm kin}$ and $\omega_{\rm spin}$ being formally 
$\Or(1/\mb)$.
At leading order (static limit), where the heavy quark acts only as a 
static colour source and the light quarks are independent of the heavy 
quark's flavour and spin, the theory is expected to have $\sim 10\%$ 
precision, while this reduces to $\sim 1\%$ at $\Or(1/\mb)$ representing the 
interactions due to the motion and the spin of the heavy quark. 
As crucial advantage (e.g., over NRQCD), HQET treats the 
$1/\mb$--corrections to the static theory as space-time insertions in 
correlations functions.
For correlation functions of some multi-local fields $\op{}$ and up to 
$1/m_{\rm b}$--corrections to the operator itself (irrelevant when spectral 
quantities are considered), this means
\beq
\langle\op{}\rangle=
\langle\op{}\rangle_{\mrm{stat}}+a^4\sum_x\left\{
\omkin\langle\op{}\Okin(x)\rangle_{\mrm{stat}}
+\omspin\langle\op{}\Ospin(x)\rangle_{\mrm{stat}}
\right\}\,,
\eeq
where $\langle\op{}\rangle_{\rm stat}$ denotes the expectation value in the 
static approximation and $\Okin$ and $\Ospin$ are given by 
$\heavyb\vecD^2\heavy$ and $\heavyb\vecsigma\vecB\heavy$.
In this way, HQET at a given order is (power-counting) renormalizable and
its continuum limit well defined, once the mass counterterm $\dmstat$ and
the coefficients $\omkin$ and $\omspin$ are fixed non-perturbatively by a 
matching to QCD.

Still, for lattice HQET and its numerical applications to lead to precise 
results with controlled systematic errors in practice, two shortcomings had 
to be left behind first.

1.) The exponential growth of the noise-to-signal ratio in static-light 
correlators, which is overcome by a clever modification of the Eichten-Hill 
discretization of the static action~\cite{HQET:statprec}.

2.) As in HQET mixings among operators~of different dimensions occur, the
power-divergent additive mass renormalization $\dmstat\sim g_0^2/a$
already affects its leading order.
Unless HQET is renormalized non-perturbatively~\cite{Maiani:1992az}, this 
divergence --- and further ones $\sim g_0^2/a^{2}$ arising at 
$\Or(1/\mb)$ --- imply that the continuum limit does not exist owing to a 
remainder, which, at any finite perturbative 
order~\cite{mbstat:dm_MaSa,mbstat:dm_DirScor}, diverges as $a\to 0$.
A general solution to this theoretically serious problem was worked out and 
implemented for a determination of the b-quark's mass in the static and 
quenched approximations as a test case~\cite{HQET:pap1}.
It is based on a \emph{non-perturbative matching of HQET and QCD in 
finite volume}.

\section{Application: The b-quark mass from HQET at $\Or(1/\mb)$}
\label{Sec_appl}
Let us first note~\cite{reviews:NPRrainer_nara} that in order not to spoil 
the asymptotic convergence of the series, the matching must be done 
non-perturbatively --- at least for the leading, static piece --- as soon as 
the $1/\mb$--corrections are included, since
as $\mb\to\infty$ the \emph{perturbative} truncation error from the 
matching coefficient of the static term becomes much larger than the power 
corrections $\sim\lQCD/\mb$ of the HQET expansion.

%
\begin{figure}[htb]
\begin{center}
\vspace{-0.5cm}
\includegraphics[width=\textwidth]{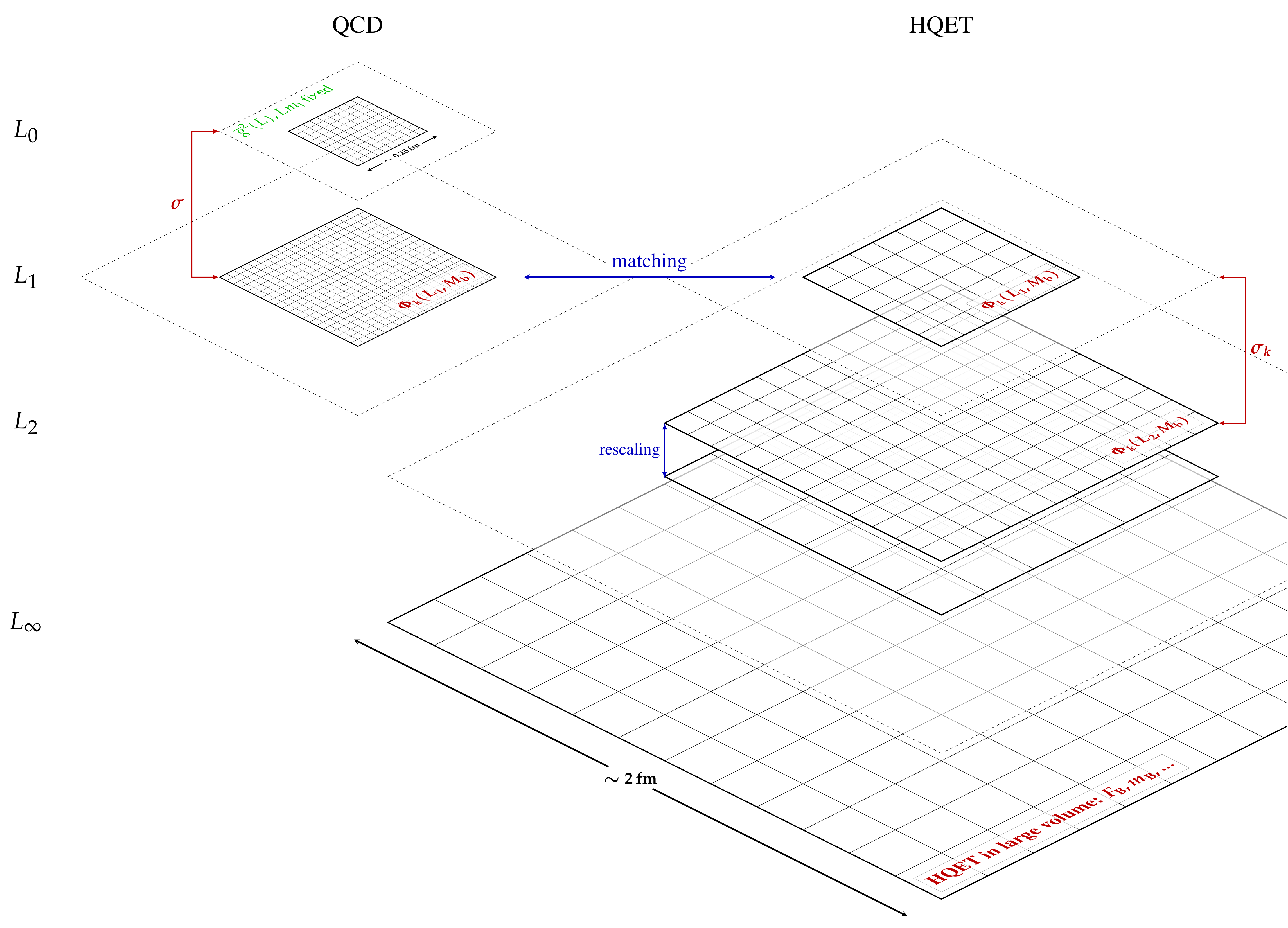}
\vspace{-0.875cm}
\caption{%
Idea of lattice HQET computations via a non-perturbative determination of 
HQET parameters from small-volume QCD simulations.
For each fixed $L_i$, the steps are repeated at smaller $a$ to reach a 
continuum limit.
}\label{fig:strat}
\end{center}
\vspace{-0.5cm}
\end{figure}
%
In the framework introduced in~\cite{HQET:pap1}, matching and 
renormalization are performed simultaneously \emph{and} non-perturbatively.
The general strategy, illustrated in~\Fig{fig:strat}, can be explained as
follows.
Starting from a finite volume with $L_1\approx 0.5\,\Fm$, one chooses 
lattice spacings $a$ sufficiently smaller than $1/\mb$ such that the 
b-quark propagates correctly up to controllable discretization errors of 
order $a^2$. 
The relation between the renormalization group invariant (RGI) and the bare 
mass in QCD being known, suitable finite-volume observables 
$\Phi_k(L_1,M_{\rm h})$ can be calculated as a function of the RGI heavy 
quark mass, $M_{\rm h}$, and extrapolated to the continuum limit. 
Next, the power-divergent subtractions are performed non-perturbatively by 
a set of matching conditions, in which the results obtained for $\Phi_k$ 
are equated to their representation in HQET (r.h.s.~of~\Fig{fig:strat}).
At the same physical value of $L_1$ but for resolutions $L_1/a=\rmO(10)$, 
the previously computed heavy-quark mass dependence of 
$\Phi_k(L_1,M_{\rm h}) $ in finite-volume QCD may be exploited to determine 
the bare parameters of HQET for $a\approx(0.025-0.05)\,\Fm$.
To evolve the HQET observables to large volumes, where contact with some 
physical input from experiment can be made, one also computes them at these 
lattice spacings in a larger volume, $L_2=2L_1$.
The resulting relation between $\Phi_k(L_1)$ and $\Phi_k(L_2)$ is encoded in
associated step scaling functions (SSFs) $\sigma_k$, indicated 
in~\Fig{fig:strat}.
By using the knowledge of $\Phi_k(L_2,M_{\rm h})$ one fixes the bare 
parameters of the effective theory for $a\approx(0.05-0.1)\,\Fm$ so that a 
connection to lattice spacings is established, where large-volume 
observables, such as the B-meson mass or decay constant, can be calculated 
(bottom of~\Fig{fig:strat}). 
This sequence of steps yields an expression of $\mB$, the physical input, as 
a function of $M_{\rm h}$ via the quark mass dependence of 
$\Phi_k(L_1,M_{\rm h})$, which eventually is inverted to arrive at the 
desired value of the RGI b-mass within HQET.
The whole construction is such that the continuum limit can be taken for 
all pieces. 
\subsection{Review of the quenched computation of the b-quark mass~\cite{HQET:mb1m}}
\label{Sec_appl_Nf0}
To apply this to $\Mb$, the task is to fix $\dmstat$ and $\omkin$ 
non-perturbatively by performing a matching to QCD, after restricting to 
spin-averaged quantities to get rid of the contributions proportional to 
$\omspin$, 
For sensible definitions of the required matching observables, 
$\Phi_1$ and $\Phi_2$, we work with the Schr\"odinger functional (SF), 
i.e., QCD with Dirichlet boundary conditions in time and periodic ones in 
space (up to a phase $\theta$ for the fermions):
$\Phi_1^{\rm QCD}(L,\mh)$ exploits the sensitivity of SF correlation functions 
to $\theta$ and $\Phi_2^{\rm QCD}(L,\mh)\equiv L\Gamma_1(L,\mh)$, where 
$\Gamma_1$ is a finite-volume effective energy. 
When expanded in HQET\footnote{%
Here, $\dmstat=0$ in the action; its effect is accounted for in the overall 
energy shift $\mhbare$ in HQET versus QCD.
}, 
$\Phi_1^{\rm HQET}(L)$ is given by $\omkin$ times a quantity defined in the 
effective theory (called $R_1^{\rm kin}(L,\theta,\theta')$), whereas
$\Phi_2^{\rm HQET}(L)$ is a function of $\omkin$ and $\mhbare=\dmstat+\mh$ 
involving two other HQET quantities, $\Gamma_1^{\rm stat}(L)$ and 
$\Gamma_1^{\rm kin}(L)$. 
According to the strategy sketched above, by equating 
$\Phi_k^{\rm QCD}(L_1,\mh)$ and $\Phi_k^{\rm HQET}(L_1)$ one can determine the 
bare parameters $\mhbare$ and $\omkin$ as functions of $\mh$ at the lattice 
spacings belonging to the volume $L_1^4$. 
To employ the spin-averaged B-meson mass, $\mB^{\rm av}$, as phenomenological
input, the $\Phi_k$ are evolved to larger volumes through proper
SSFs, where the resulting $\Phi_k^{\rm HQET}(2L_1,\mh)$ still carry the 
dependence on $\mh$ inherited from the matching to QCD in $L_1^4$.
After 2 evolution steps (and taking continuum limits), linear extents of 
$\gtrsim 1.5\,\Fm$ are reached, and $\mhbare$ and $\omkin$, expressed in terms 
of SSFs, $\Phi_k^{\rm QCD}(L_1,\mh)$ as well as $R_1^{\rm kin}$, 
$\Gamma_1^{\rm stat}$ and $\Gamma_1^{\rm kin}$, are obtained 
--- again as functions of $\mh$.
Now, the b-quark mass is extracted by solving
\beq
\mB^{\rm av}= 
E^{\rm stat}+\omkin(\mh)E^{\rm kin}+\mhbare(\mh)
\label{mastereq_E}
\eeq
for $\mh$, with 
$E^{\rm stat}=\lim_{L\to\infty}\Gamma_1^{\rm stat}$ and
$E^{\rm kin}=-\langle\,
{\rm B}\,|\,a^3\sum_{\bf z}\Okin(0,\fat{z})\,|\,{\rm B}\,\rangle_{\rm stat}$.
All quantities entering \eq{mastereq_E} have a continuum limit either in 
QCD or HQET, which implies that all power divergences have been subtracted 
non-perturbatively.

%
\begin{figure}[htb]
\begin{center}
\vspace{-3.125cm}
\hspace{0.25cm}
\includegraphics[width=0.625\textwidth]{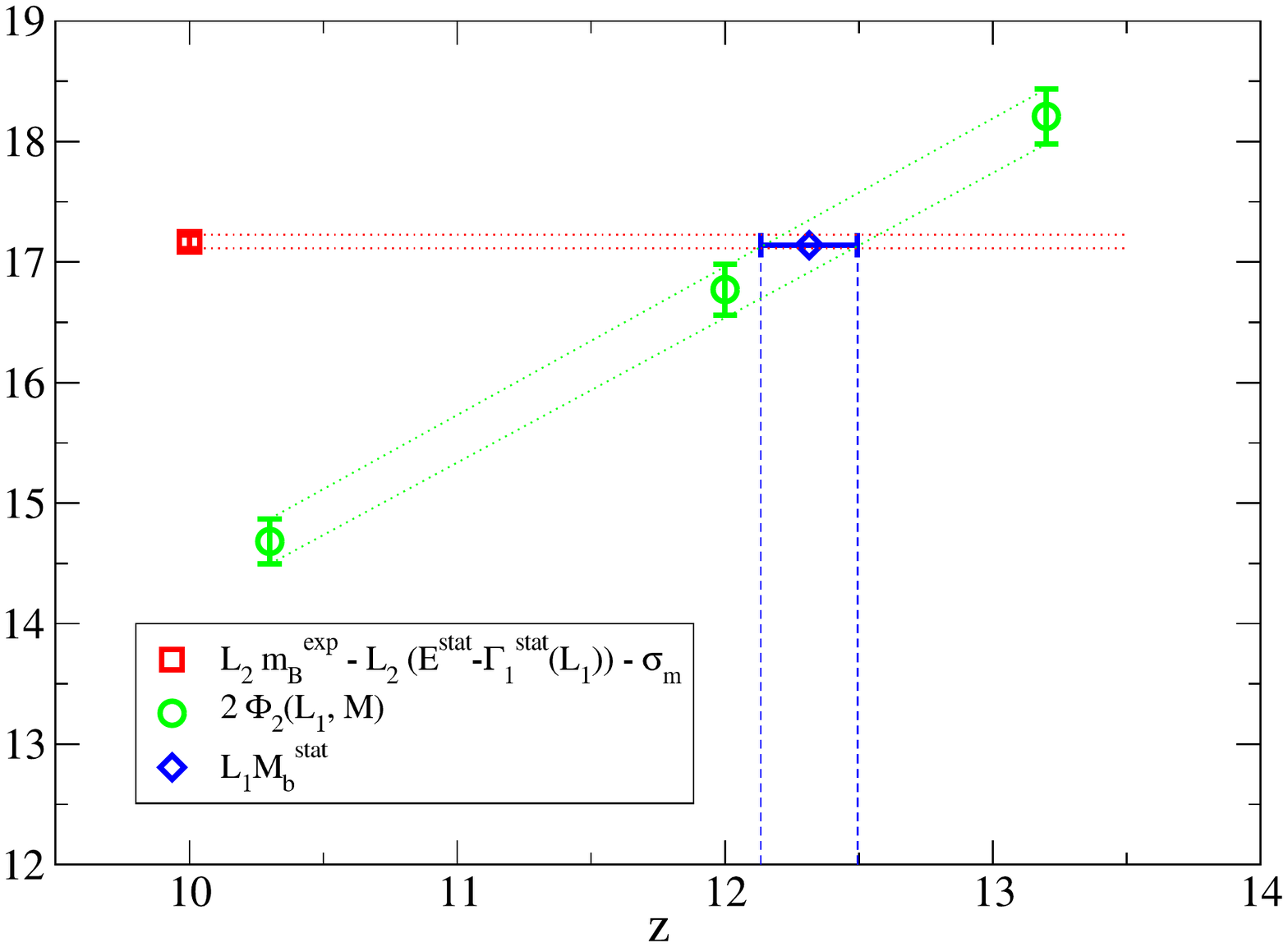}
\vspace{-3.25cm}
\caption{%
Graphical solution of \eq{mastereq_diff} in the quenched 
case~\cite{HQET:mb1m}.
The quantity used in the finite-volume matching step is 
$\Phi_2^{\rm QCD}(L_1,M)=L_1\Gamma_1(L_1,M)$, where $z=L_1M$ and $M\equiv\Mh$
the RGI heavy quark mass.
}\label{fig:Mb_stat}
\vspace{-0.5cm}
\end{center}
\end{figure}
%
In case of the leading-order, static approximation, where only $\mhbare$ 
needs to be determined, the small- and large-volume matching conditions
simplify to $\Gamma_1(L_1,\mh)=\Gamma_1^{\rm stat}(L_1)+\mhbare$ and 
$\mB^{\rm av}=E^{\rm stat}+\mhbare$, respectively.
To be able to solve the first equation for  $\mhbare$ and replace it 
in the second, we bridge the volume gap in two steps by inserting a SSF 
$\sigm(L_1)=2L_1[\Gamma_1^{\rm stat}(2L_1)-\Gamma_1^{\rm stat}(L_1)]$ and arrive
at the \emph{master equation}
\beq
L_1\left[\,\mB^{\rm av}-(E^{\rm stat}-\Gamma_1^{\rm stat})\,\right]
-{\T \frac{\sigm(L_1)}{2}}=L_1\Gamma_1(L_1,\mh)\,,
\label{mastereq_diff}
\eeq
where $\Gamma_1$ originates from QCD in $L_1^4$ and any reference to bare 
parameters has disappeared.
Its graphical solution is reproduced in~\Fig{fig:Mb_stat} and yields
$\Mb^{\rm stat}=6.806(79)\,\GeV$.

The inclusion of the sub-leading $1/\mb$--effects is technically more 
involved and exploits the freedom of choices for the angle(s) $\theta$ and 
an alternative set of matching observables~\cite{HQET:mb1m}. 
We just quote the final value $\mbbMS(\mbbar)=4.347(48)\,\GeV$ with the 
remark that, upon including the $1/\mb$--terms, differences among the 
static results w.r.t.~the matching condition chosen are gone, which 
signals practically negligible higher-order corrections.
\subsection{Status in two-flavour QCD}
\label{Sec_appl_Nf2}
The renormalization of HQET through the non-perturbative matching to 
$N_{\rm f}=2$ QCD in finite volume, to do the power-divergent subtractions, 
is under way~\cite{lat07:hqetNf2,lat08:hqettests,HQET:Nf2}.
As an important prerequisite, the calculated non-perturbative relation 
between the RGI and subtracted bare heavy quark mass \cite{impr:babp_nf2} 
enables to fix RGI heavy quark masses in the matching volume $L_1^4$:
\beq
L_1M=
\zM(g_0)Z(g_0)\left(1+\bm(g_0)a\mqh\right)L_1\mqh\,, \quad M=\Mh\,.
\eeq
The extent $L_1$ is defined via a constant SF coupling, 
$\gbsq(L_1/2)=2.989$, and the PCAC masses of the dynamical light quarks are 
tuned to zero.

\begin{figure}[htb]
\begin{center}
\vspace{-0.375cm}
\includegraphics[width=0.495\textwidth]{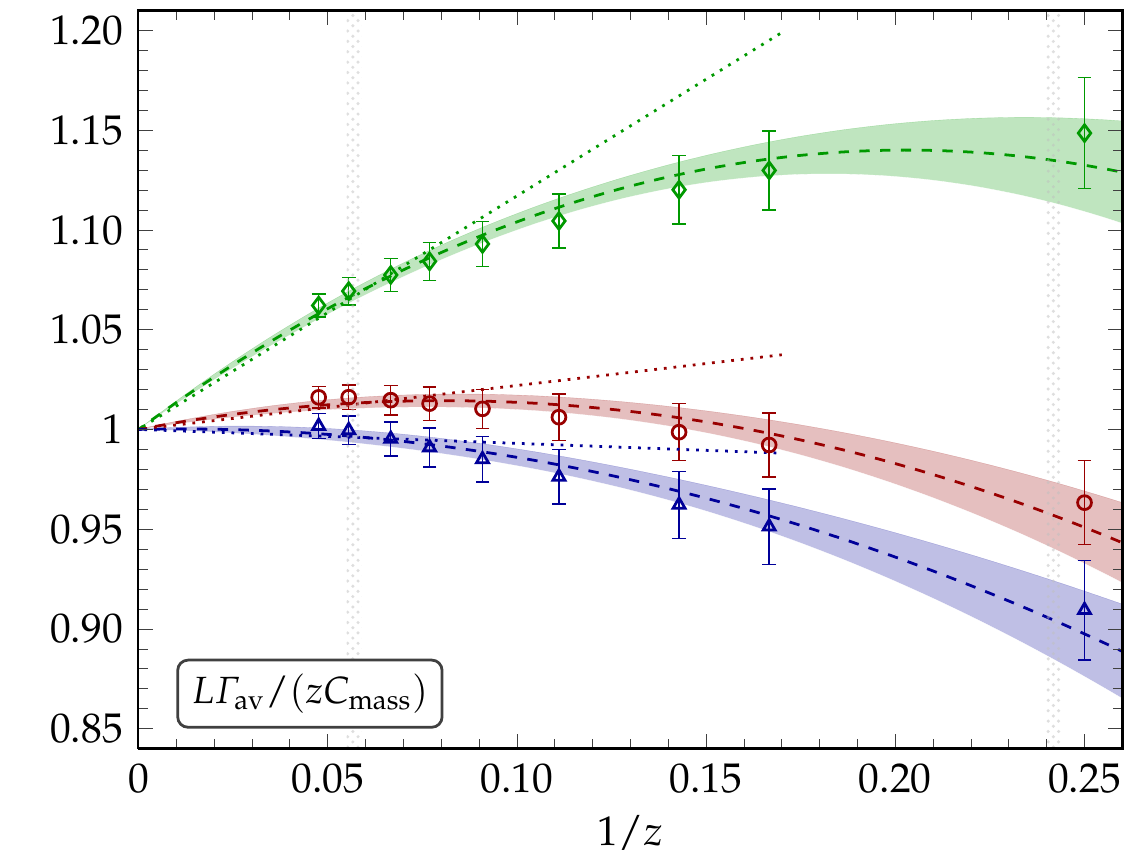}
\includegraphics[width=0.495\textwidth]{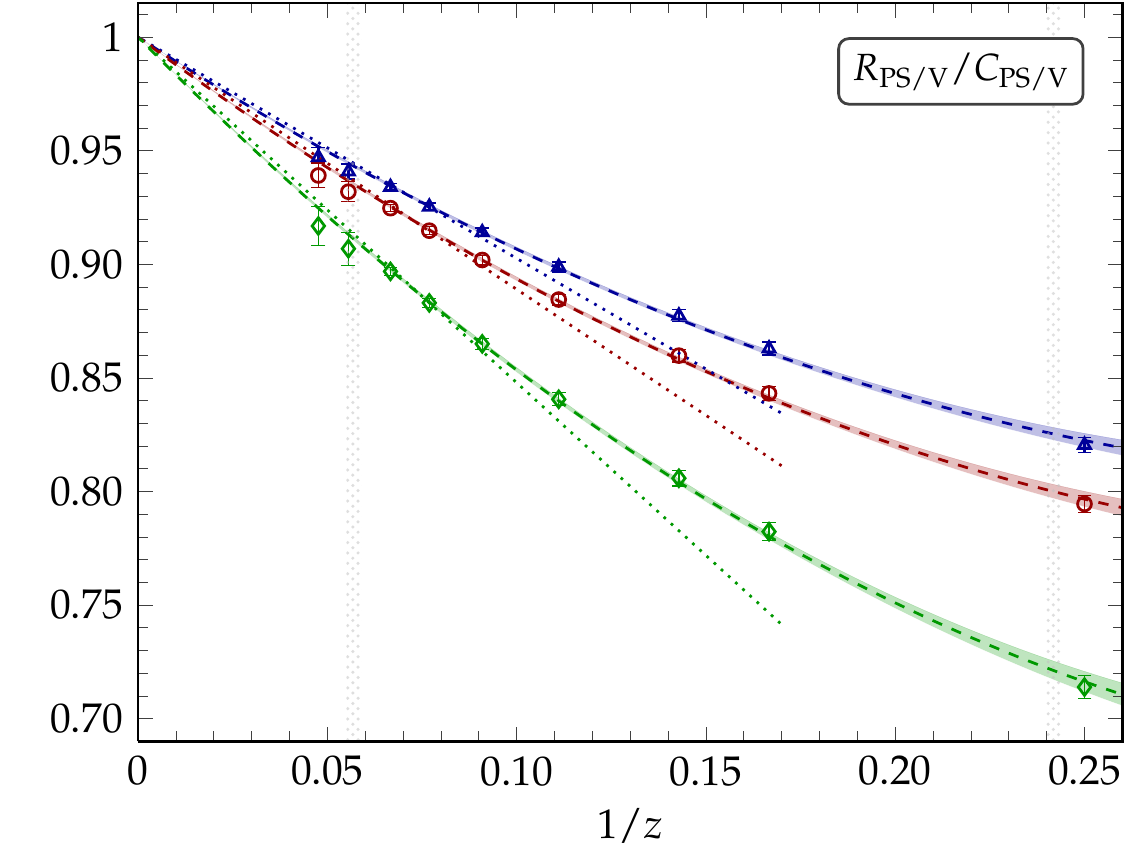}
\vspace{-0.825cm}
\caption{%
$z$--dependence of two QCD observables, which enter the matching of QCD 
with HQET in $\nf=2$ with massless dynamical quarks.
$z=L_1M$ as before.
Left: Spin-averaged B-meson energy.
Right: Ratio of axial-vector to vector current matrix elements.
The conversion functions $C_{\rm X}$, which only introduce small perturbative 
uncertainties \cite{lat08:hqettests,HQET:Nf2}, translate the HQET 
predictions to the corresponding QCD quantities at finite values of $z$. 
}\label{fig:l1zdep}
\vspace{-0.5cm}
\end{center}
\end{figure}

\Fig{fig:l1zdep} shows two examples for the heavy quark mass dependence of 
finite-volume QCD observables in the continuum limit, which enter the 
non-perturbative matching.
%
%

These results also allow to perform non-perturbative tests of HQET in the 
spirit of the corresponding quenched 
investigation~\cite{HQET:pap3}.
The calculation of the step scaling functions in HQET is expected to be 
finished soon, and the large-volume part of our strategy is currently
being implemented within the CLS effort~\cite{community:CLS}.

\section{Outlook}
\label{Sec_outl}
The non-perturbative treatment of HQET including $1/\mb$--terms can lead
to results with unprecedented precision for B-physics on the lattice. 
It also greatly improves our confidence in the use of the effective theory.
The striking agreement, for example, between the decay constant 
$\Fbs$ computed including $1/\mb$--corrections and the value 
resulting from the interpolation between the static number and data around 
the charm \cite{fbstat:pap2,HQET:Nf0}, though still in the quenched 
approximation, provides a strong internal check of the approach.
In addition, the HQET parameters at $\Or(1/\mb)$ calculated 
non-perturbatively by the ALPHA Collaboration \cite{HQET:Nf0} can be 
employed for several other quantities.
The programme aiming to reach the same accuracy in the $\nf=2$ dynamical
case is well advanced and progressing fast~\cite{HQET:Nf2}.

\vspace{0.125cm}
\noindent {\bf Acknowledgments.}
We are indebted to our colleagues in CLS and ALPHA for a fruitful 
collaboration, and to P.~Fritzsch for a part of the figures.
We acknowledge support by the Deutsche Forschungsgemeinschaft in the 
SFB/TR~09-03, ``Computational Particle Physics'', and under grant
HE~4517/2-1, as well as by the European Community through EU Contract 
No.~MRTN-CT-2006-035482, ``FLAVIAnet''.
Our simulations are performed on BlueGene and APE Machines of the John von 
Neumann Institute for Computing at Forschungszentrum J\"ulich, at DESY, 
Zeuthen, and INFN, University of Rome ``Tor Vergata''.
We also thankfully acknowledge the computer resources and technical support
for the CLS simulations provided by the HLRN in Berlin, 
the Universities of Mainz, Rome ``La Sapienza'' and Valencia-IFIC, by CERN, 
and by the Barcelona Supercomputing Center.

\bibliographystyle{nic}
\vspace{-0.25cm}
\bibliography{lattice_ALPHA}
\end{document}